\documentclass[twocolumn,showpacs,preprintnumbers,amsmath,amssymb]{revtex4}
\usepackage{graphicx}
\usepackage{dcolumn}
\usepackage{bm}

\begin{document}
\title{Statistical fluctuations of cooperative radiation produced by nonisochronous electrons-oscillators}

\author{S.V.~Anishchenko}
\email{sanishchenko@mail.ru}
\affiliation{Research Institute for Nuclear Problems\\
Bobruiskaya str., 11, 220030, Minsk, Belarus.}%

\author{V.G.~Baryshevsky}
 \email{bar@inp.bsu.by}
 \affiliation{Research Institute for Nuclear Problems\\
Bobruiskaya str., 11, 220030, Minsk, Belarus.}%

\begin{abstract}
Shot noise, intrinsic to the ensemble of nonisochronous
electrons-oscillators, is the cause of statistical fluctuations in
cooperative radiation generated by single-pass cyclotron-resonance
masers (CRMs). Autophasing time --- the time required for the
cooperative radiation power to peak --- is the critical parameter
characterizing the dynamics of electrons-oscillators interacting
via the radiation field. Shot-noise related fluctuations of the autophasing time imposes appreciable
limitations on the possibility of coherent summation  of
electromagnetic oscillations from several single-pass CRMs.

Premodulation of charged particles leads to a considerable
narrowing of the autophasing time distribution function. When the
number of particles $N_e$ exceeds a certain value that
depends on the degree to which the particles have been
premodulated, the relative root-mean-square deviation (RMSD)
of the autophasing time $\delta_T$ changes from a logarithmic 
dependence on $N_e$ ($\delta_T\sim1/\ln
N_e$) to  square-root ($\delta_T\sim1/\sqrt{N_e}$). As a result, there is an increased probability of coherent
summation of electromagnetic oscillations from several single-pass
generators.

A slight energy spread ~($\sim$4\%) results in a twofold drop of the
maximum attainable  power of cooperative radiation.

\end{abstract}

\pacs{41.60.-m, 05.40.Ca, 84.40.lk}

\maketitle

\section{Introduction}
R. H. Dicke in his pioneering  work \cite{Dicke1954} has shown
that a system of $N_a$ number of inverted  two-level atoms can
spontaneously transit to the ground state during the time
inversely proportional to the number of atoms $N_a^{-1}$. Because
the total energy emitted  by the collection of atoms is
proportional to $N_a$, the radiation intensity reveals a
square dependence on $N_a$. This type of radiation was termed
the "collective spontaneous radiation", or the "Dicke
superradiance". A remarkable progress has been achieved in the
understanding of this phenomenon in~\cite{Bonifacio1971}. The
authors of~\cite{Bonifacio1971} have developed a quantum theory of
superradiance for a single-mode model. The simplicity of the model
enables a detailed study of the  kinetics and statistical
properties of superradiance. Later, it has been found \cite{DeGiorgio1971,Gross1982,Andreev1988}
that the radiation intensity undergoes appreciable statistical
fluctuations and the RMSD of the superradiative
instability time $T$ slowly decreases according to a logarithmic
law as the number of particles is increased:
$\delta_T\approx1.3/\ln N_a$.


The elementary unit of the system we have just described is a
two-level atom, the concept  widely used in quantum
electronics and optics to describe physical processes. In the
excited state, the two-level atom interacting with the radiation
field is also involved in relaxation  processes caused by
inter-particle collisions.
If the radiation growth time $T$ is much greater
than the relaxation time, then the pulse of superradiance  is not
formed, and the particles emit incoherently. This greatly restricts
the selection of  active medium  for experimental
investigation of superradiance and design of superradiance-based
short-pulse electromagnetic sources
\cite{Skribanowitz1973,MacGillivray1976}.

In this connection, a considerable attention has been
paid recently to the generation of short superradiance pulses
\cite{Bonifacio1990,Ginzburg2013}, also
termed as "self-amplified spontaneous emission", "superradiance", or "cooperative radiation"
\cite{Trubeckov2005},
using short electron beams propagating in complex electrodynamical
structures (undulators, resonators, dielectric and corrugated
waveguides, or photonic crystals)
\cite{Bonifacio1990,Ginzburg2013,AnishchenkoBaryshevsky2015,Rostov2012,Ginzburg2015}.
In such systems, the average relaxation time that depends on the
collisions between charged particles exceeds manifold the  time
needed for the formation of the cooperative-radiation pulse.

Cooperative radiation is generated via a nonlinear interaction of
charged particles and the electromagnetic field, causing the
electrons to bunch. As a result, the whole of the  electron beam
becomes spatially modulated.  Spatial
modulation is accompanied by a coherent emission of
electromagnetic waves. 


%

The possibility to produce short pulses of cooperative radiation
was first substantiated in \cite{Kondratenko1980,Bonifacio1984}
for free electron lasers (FELs). However, the
experimentally measured output characteristics of  cooperative
radiation (power, radiation spectra, and energy in a pulse) seemed
really stochastic, which, according to theoretical
\cite{Bonifacio1994,Saldin1998} and experimental
\cite{Hogan1998,Andruszkow2000,Yurkov2002,Ayvazyan2003,Lehmkuhler2014}
studies, was closely related to  the presence of the shot noise
inherent in charged-particle beams. It is noteworthy that high
 frequency and time stability of the radiation pulse
 is crucial for many applications. For this reason, electron
 beams are premodulated at the radiation frequency to reduce
 the influence of shot noise on cooperative radiation \cite{Yu1991,Yu2000,Allaria2012,Amann2012,Ninno2015}.

In a microwave range, single-pass CRMs
\cite{Ginzburg1994,Ginzburg1996,Ginzburg1997a} and Cherenkov
generators are more commonly used
\cite{Ginzburg1997b,McNeil1999,Wiggins2000}. The generators of
cooperative radiation operate in two regimes: traveling wave
\cite{Ginzburg1997b,McNeil1999,Wiggins2000} and backward wave
\cite{Ginzburg1999,Ginzburg2002,Elchaninov2003a,Elchaninov2003b,Korovin2006}.

The distinguishing  feature of the  traveling-wave regime is that
the group velocity of electromagnetic waves and the velocity of
charged particles are co-directional. This regime was applied in
first experiments with single-pass CRMs
\cite{Ginzburg1996,Ginzburg1997a} and Cherenkov generators
\cite{Ginzburg1997b}.
 Theoretical consideration of radiation gain in short-pulse traveling-wave generators revealed an
 important peculiarity: the stability of the cooperative radiation
 parameters can be  improved noticeably by injecting into the electrodynamical structure of
a beam with a sharp front whose duration is comparable with the emission
period \cite{McNeil1999,Wiggins2000}.  In this case, the Fourier
transform of the beam current contains quite a significant
spectral component at the radiation frequency. As a result, the
generation of electromagnetic oscillations starts with coherent
spontaneous emission of the whole beam instead of  incoherent
spontaneous emission from individual particles. As a result, the
degree of fluctuations in cooperative radiation is decreased.

Though the early single-pass CRMs and Cherenkov generators
operated in the traveling-wave  regime, the most impressive
results were obtained in experiments with single-pass
backward-wave tubes  generating cooperative radiation whose peak
power was
 appreciably greater than  the beam power
\cite{Elchaninov2003a,Elchaninov2003b,Korovin2006}.
For example, the  peak radiated power of 1.2 GW was attained at
$9.3$ GHz for beam power of 0.87 GW \cite{Elchaninov2003a}. The
stability of output parameters of the cooperative-radiation pulse
generated in the backward-wave regime, as in a single-pass traveling-wave generator, strongly
depended on the
 beam front.

The analysis of theoretical and experimental works on generation
of cooperative radiation reveals that the main cause of
statistical spread of the output characteristics is
the shot noise  inherent in the electron beam.
The impact it produces can be diminished by premodulating the
beam at the radiation frequency
\cite{Yu1991,Yu2000,Allaria2012,Amann2012,Ninno2015}.
Naturally, one wonders  what degree of beam premodulation is required
for certain applications.
This problem would be most prevalent for high-current
generators with electron beams composed of ectons, individual
portions of electrons that  contain up to $10^{11}$ elementary
charge carriers \cite{Mesyats1975,Mesyats2005}.
Taking into account the  complex structure of the electron beam is
a must for  accurate estimations of the shot noise
\cite{Abubakirov2009}, which is
essential in solving the problem of coherent summation of
electromagnetic oscillations from several short-pulse sources of
radiation \cite{Rostov2012,Ginzburg2015}.

This paper considers the effect of shot noise on the generation of
cooperative radiation from a premodulated short beam of
particles.
We shall consider this issue by the example of an ensemble of
nonisochronous electrons-oscillators interacting with one another
via the  radiation field, used  as one of the basic models
for the description of nonlinear generation of electromagnetic
waves in CRMs \cite{Gaponov1967,Zheleznjakov1986,Ilyinsky1988,Wainshtein1990,Kobelev1991,AnishchenkoBaryshevsky2016}.
It has been shown in \cite{Ilyinsky1988,Wainshtein1990} that in
the absence of external action,  the instability  evolves in the
ensemble of nonisochronous electrons-oscillators, accompanied at
the initial stage  by an exponential growth of the radiated power
and autophasing of electrons-oscillators. This exponential growth
is then suppressed  due to nonlinearity,  and the pulse of
cooperative radiation is formed
 \cite{Ilyinsky1988}.
 The influence  of nonisochronism on the peak power of cooperative
radiation and the  autophasing time of electrons-oscillators was
studied in detail by Vainstein and colleagues
 \cite{Wainshtein1990}. However, the authors of \cite{Wainshtein1990}  assumed
 that the beam  had been premodulated at the radiation frequency and left out the effects related
 to  shot noise \cite{AnishchenkoBaryshevsky2016}.
 %


In our analysis  of  statistical properties of cooperative
radiation, the peak radiated power and autophasing time serve as
random variables.  The peak power is the major output
characteristic of  short-pulse electromagnetic sources, and the
autophasing time is the parameter defining the minimum time of
particle passage through the generator needed for cooperative
instability to evolve.  Moreover, the RMSD in the autophasing time is of fundamental importance for
coherent summation of oscillations from several
cooperative-radiation sources.


The paper's outline is as follows. First, we derive a system of
equations describing the interaction of nonisochronous oscillators
via the radiation field by the example of electron ensemble
circulating in a uniform magnetic field.
Further comes a detailed consideration of statistical fluctuations
 of cooperative radiation in the presence of shot noise
from the ensemble of nonisochronous electrons-oscillators with and
without phase premodulation of charged particles. Particular attention is given to finding the
autophasing time distribution function.
Based on the described study, we shall then investigate in detail the
statistical fluctuations in single-pass CRMs and discuss the
limits imposed on the output characteristics thereof by the shot noise and the
energy spread. Furthermore, we
shall discuss the possibility of coherent summation of
oscillations coming from several single-pass sources of  cooperative cyclotron
radiation.

\section{Cooperative cyclotron radiation}

Let us consider the behavior of a nonrelativistic electron beam in
a uniform magnetic field~$\vec H$ directed along the  $OZ$ axis in
the presence of the radiative energy loss. (The behavior of
the relativistic electron beam is discussed in Appendix A.)
Particle velocity components perpendicular and parallel to the
magnetic-filed vector are denoted by~$\vec v_{\perp k}$ and $\vec
v_{z k}$, respectively.  Then the behavior of particles is
described by the equations of motion in the form
\begin{equation}
\label{Reaction1}
\begin{split}
&\dot {\vec p}_{\perp k}=\frac{e}{c}\vec v_{\perp k}\times\vec H+\vec F_{\perp k},\\
&\dot p_{zk}=F_{zk}.\\
\end{split}
\end{equation}
Here,  $\vec F_k$ is the force acting on the $k$th particle from
all particles and  $\vec p_k$ is its  momentum  related to the
velocity~$\vec v_k$ as
\begin{equation}
\label{oscillators2}
\vec p_k=\frac{m\vec v_k}{\sqrt{1-v_k^2/c^2}}\approx m\vec v_k\Big(1+\frac{v_{\perp k}^2+v_{zk}^2}{2c^2}\Big).
\end{equation}

If the beam size is less than the radiation
wavelength $\lambda=2\pi mc/eH$, and the Coulomb repulsion force
and induction fields can be neglected (see Appendix B), then the
force acting on each particle is given by the sum (see Appendix C)
\begin{equation}
\label{Reaction2}
\vec F_k=e\sum_j\frac{2e}{3c^3}\ddot{\vec v}_j,
\end{equation}
where  $\frac{2e}{3c^3}\ddot{\vec v}_j$ is the radiation
field induced by the $j$th particle.

Let us pay attention to an essential circumstance \cite{Landau2}:
the expression for  $\vec F_k$ is true if the radiative friction
force is appreciably less than the Lorentz force $\frac{e}{c}\vec
v_k\times\vec H$ acting on each particle. Otherwise, unphysical
self-accelerated solutions may arise.

The requirement that the radiative friction force should be much
less than the Lorentz force imposes limitation on the magnetic
field strength (in the opposite case the considered theory is invalid). For single particle \cite{Landau2}:
\begin{equation}
\label{Reaction3}
 H\ll\frac{m^2c^4}{e^3}\approx6\cdot10^{15}\text{ Gs}.
\end{equation}
But in the case of a dense beam of coherently emitting particle with  $N_e$
electrons,  mass  $M=N_em$, and charge  $Q=N_ee$  we can write by
analogy with~\eqref{Reaction3}
\begin{equation}
\label{Reaction4}
 H\ll H_{cr}=\frac{M^2c^4}{Q^3}=\frac{m^2c^4}{N_ee^3}.
\end{equation}
With the present-day  acceleration facilities, dense beams with
$N_e\sim10^{10}$ electrons are available; substitution of
$N_e\sim10^{10}$ into  \eqref{Reaction4} gives  $H\ll 60$~kGs. If
this condition is violated, the equation set \eqref{Reaction1}
with the force \eqref{Reaction2} is inapplicable.

Thus, if the radiation reaction force is much less than the
Lorentz force and the beam size is less than the radiation
wavelength, then the equations of motion describing the
interaction between charged particles have the form:
\begin{equation}
\label{oscillators1}
\begin{split}
&\dot p_{\perp k}=\frac{e}{c}\vec v_k\times\vec H+\frac{2e^2N_e}{3c^3}\ddot{\vec v}_{\perp},\\
&\ddot{\vec v}_\perp=\frac{1}{N_e}\sum_k\ddot{\vec v}_{\perp k}.\\
\end{split}
\end{equation}

In the absence of energy losses through emission, the particles
are in circular motion with cyclic frequencies  \cite{Landau2}
\begin{equation}
\label{oscillators3}
\Omega_k=\frac{eH}{mc}\sqrt{1-\frac{v_k^2}{c^2}}\approx\frac{eH}{mc}\Big(1-\frac{v_{\perp k}^2+v_{zk}^2}{2c^2}\Big),
\end{equation}
depending on  $\vec v_{\perp k}$, which  is responsible for
nonisochronism of oscillations.

Using the approximate relation
\begin{equation}
\label{oscillators4}
\ddot{\vec v}_{\perp k}\approx-\Omega_k^2\vec v_{\perp k}\approx-\Omega^2\vec v_{\perp k},
\end{equation}
where
\begin{equation}
\label{oscillatots5}
\Omega=\frac{eH}{mc},
\end{equation}
we shall write vector equations \eqref{oscillators1} in a
component-wise fashion
\begin{equation}
\label{oscillators6}
\begin{split}
&\dot v_{xk}=\Omega\Big(1-\frac{1}{2}\frac{\vec v_{\perp k}^2+\vec v_{zk}^2}{c^2}\Big) v_{yk}-\frac{2e^2\Omega^2N_e}{3mc^3}v_{x},\\
&\dot v_{yk}=-\Omega\Big(1-\frac{1}{2}\frac{\vec v_k^2+\vec v_{zk}^2}{c^2}\Big) v_{xk}-\frac{2e^2\Omega^2N_e}{3mc^3}v_{y},\\
&\dot v_{zk}=0, \\
\end{split}
\end{equation}
where
\begin{equation}
\label{oscillators6b}
\begin{split}
&\vec v_x=\frac{1}{N_e}\sum_k\vec v_{x k},\\
&\vec v_y=\frac{1}{N_e}\sum_k\vec v_{y k}.\\
\end{split}
\end{equation}
Then we shall multiply the second equation  \eqref{oscillators6}
by $-i$ and  add it to the first one:
\begin{equation}
\label{oscillators6c}
\begin{split}
&\dot v_{xk}-i\dot v_{yk}=i\Omega\Big(1-\frac{1}{2}\frac{\vec v_{\perp k}^2+\vec v_{zk}^2}{c^2}\Big)(v_{xk} -iv_{yk})\\
&-\frac{2e^2\Omega^2N_e}{3mc^3}(v_{x}-iv_y).\\
\end{split}
\end{equation}

Assuming that all $v_{zk}=v_z$ are equal, let us introduce the following notation:
$a_k=e^{-i\Omega(1-v_{z}^2/2c^2)t}(v_{xk}-iv_{yk})/v_{\perp 0}$
($v_{\perp 0}=\sum_k v_{\perp k}(0)/N_e$ is the average tangential speed
of particles). Then  \eqref{oscillators6c} takes the form \cite{Ilyinsky1988}
\begin{equation}
\label{oscillators7}
\begin{split}
&\frac{da_k}{d\tau}+i\theta|a_k|^2a_k=-a,\\
&a=\frac{1}{N_e}\sum_ka_k,\\
&\theta=\frac{3mv_{\perp0}^2c}{4e^2\Omega N_e},\\
&\tau=\frac{2e^2\Omega^2N_e}{3mc^3}t.\\
\end{split}
\end{equation}
The equation set  \eqref{oscillators7} fully describes the
behavior of the ensemble of electrons-oscillators moving in a
uniform magnetic field in the presence of the radiative energy loss.

The kinetic energy  $E_{rad}$ of  transverse motion of the
oscillators, the radiation power $P_{rad}$, and the time $t$ are
related to the dimensionless quantities by
\begin{equation}
\label{Wainstein2}
E_{rad}/E_U=\frac{1}{N_e}\sum_k|a_k|^2,
\end{equation}
\begin{equation}
\label{Wainstein3}
P_{rad}/P_U=2|a|^2,
\end{equation}
and
\begin{equation}
\label{Wainstein4}
t/T_U=\tau,
\end{equation}
where
\begin{equation}
\label{Wainstein2b}
E_{U}=\frac{N_emc^2}{2}\frac{v_{\perp0}^2}{c^2},
\end{equation}
\begin{equation}
\label{Wainstein3b}
P_{U}=\frac{2e^4H^2N_e^2}{3m^2c^2}\frac{v_{\perp0}^2}{c^2},
\end{equation}
and
\begin{equation}
\label{Wainstein4b}
T_U=\frac{3m^3c^5}{2e^4H^2N_e}=\frac{2\theta c^2}{\Omega v_{\perp0}^2}.
\end{equation}

Let  $H=6.4$~kGs, $v/c=0.38$, and  $N_e=10^{9}$, then
$E_U=6~\mu$J, $P_U=0.5$~kW, $T_U=13$~ns, and  $\Omega=18$~GHz. In
this case, the nonisochronism parameter is
\begin{equation}
\theta=\frac{3m^2c^4}{4e^3HN_e}\frac{v_{\perp0}^2}{c^2}=100.
\end{equation}

Let us note that by adding the term of the form $-i\nu a_k$ to the
left-hand side of \eqref{oscillators7}, we have that under the
same initial conditions,  all complex amplitudes  $a_k$ are
multiplied by the additional phase factor $e^{i\nu \tau}$ having
no effect either on the radiation power or on the autophasing
time. This enables  analyzing different systems from a common
standpoint, and so the equation set describing the behavior of the
ensemble of weakly nonlinear electrons-oscillators oscillating in
anharmonic potential has the form \cite{Wainshtein1990}:
\begin{equation}
\label{wainshtein1}
\begin{split}
 \frac{da_k}{d\tau}+i\theta(|a_k|^2-1)a_k=-a,\\
a=\frac{1}{N_e}\sum_ka_k.\\
\end{split}
\end{equation}
Here, the nonisochronism parameter $\theta$ is selected as $\nu$.
The normalized radiation power and electron energy are given by
formulas \cite{Wainshtein1990}
\begin{equation}
P=2|a|^2
\end{equation}
and
\begin{equation}
E=\frac{1}{N_e}\sum_k|a_k|^2,
\end{equation}
respectively. In the absence of the energy spread, the absolute
values of the amplitudes  $|a_k(0)|$ equal unity at the initial
time and the phases $\phi_k(0)=\text{arg}(a_k(0))$ are uniformly
distributed in the interval  $[0;2\pi)$.

\section{Shot noise}

Shot noise in the ensemble of the electrons-oscillators is due to a
random distribution of the initial phases  $\phi_k(0)$. Each set
of  $\phi_k(0)$ corresponds to different initial conditions of the
equation set  \eqref{wainshtein1}. Fluctuations in the initial conditions
lead to fluctuations in the output characteristics of cooperative
radiation which we study by a numerical experiment.

As follows from  \eqref{wainshtein1}, the behavior of the ensemble
of nonisochronous oscillators in the absence of the energy spread
is determined by two fixed  parameters $\theta$ and $N_e$ and a
random set of initial phases $\phi_k(0)$.
Because  the distribution of initial phases $\phi_k(0)$ is random,
the numerical experiment with each pair of values of $\theta$ and
$N_e$ must have many runs. This procedure will give
information about statistical characteristics of cooperative
radiation, the most important of which are peak power $P_0$,
autophasing time $T_0$,  and their relative RMSD --- $\delta_P$ and
$\delta_T$.

\begin{figure}[ht]
\begin{center}
 \resizebox{65mm}{!}{\includegraphics{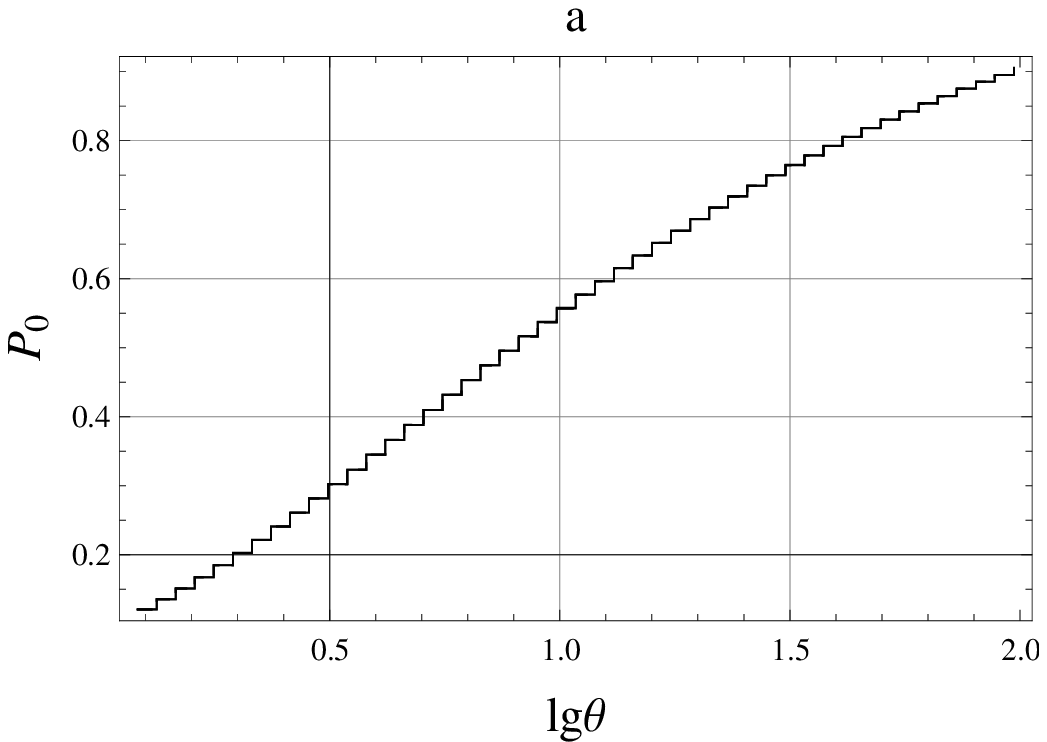}}\\
 \resizebox{65mm}{!}{\includegraphics{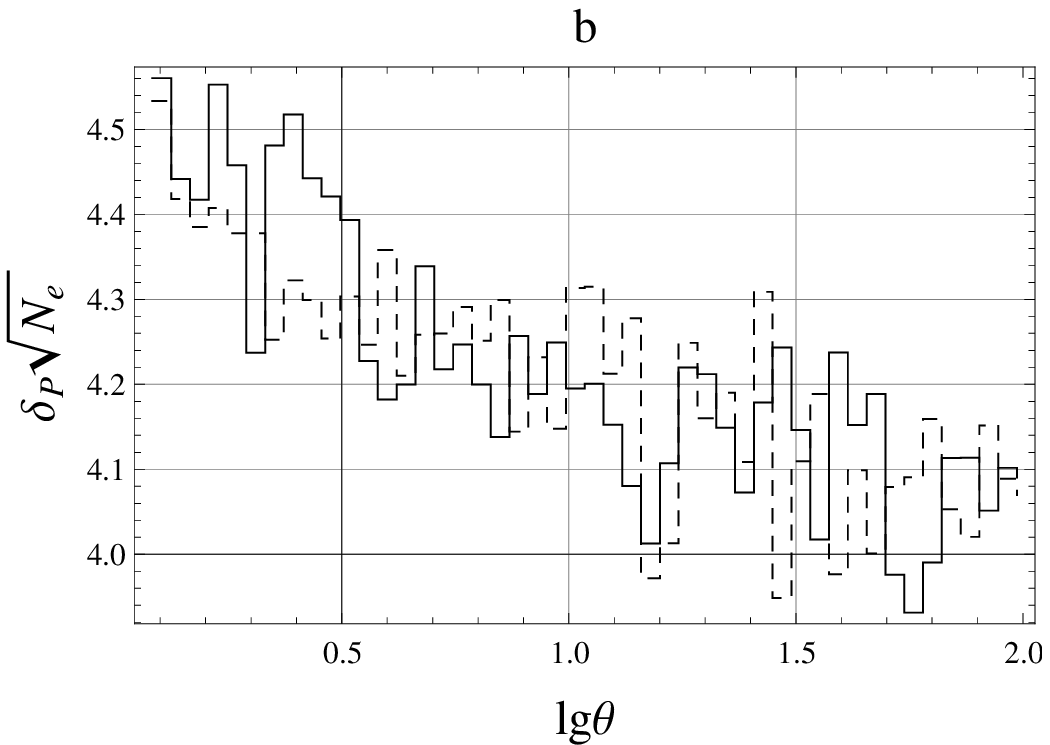}}\\
\end{center}
\caption{Peak radiated power (a) and its RMSD
(b) versus the nonisochronism parameter $\theta$
 [black curve --- $N_e=6.75\cdot10^4$ and dashed curve --- $1.08\cdot10^6$].} \label{Fig.1}
\end{figure}

\begin{figure}[ht]
\begin{center}
 \resizebox{65mm}{!}{\includegraphics{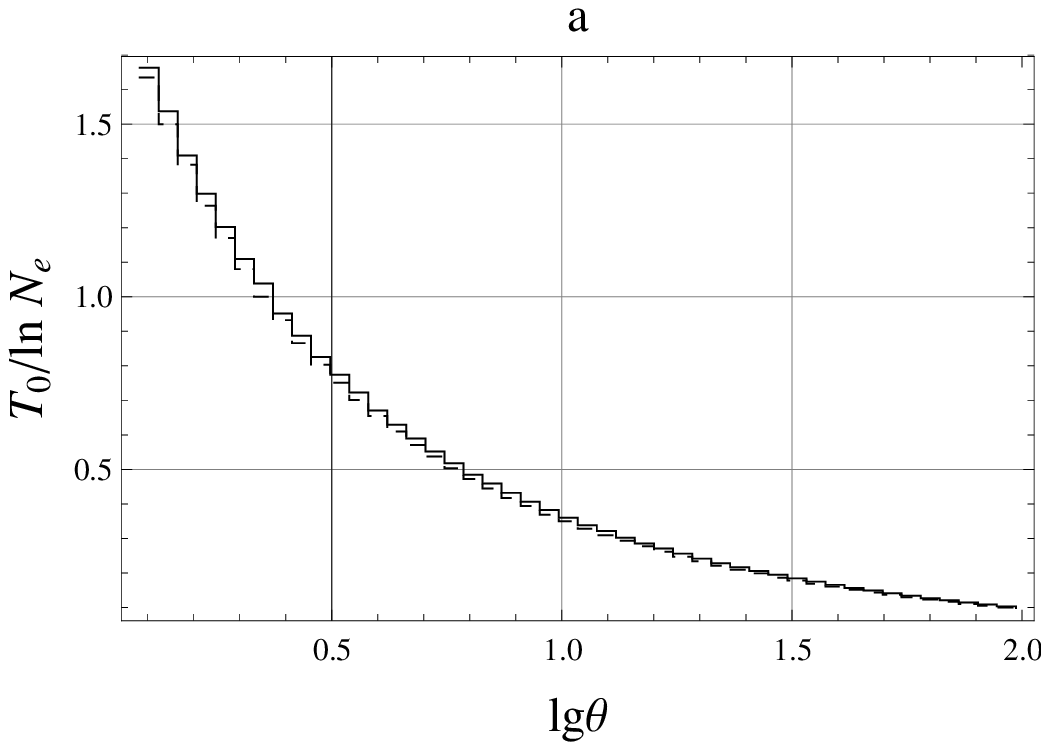}}\\
 \resizebox{65mm}{!}{\includegraphics{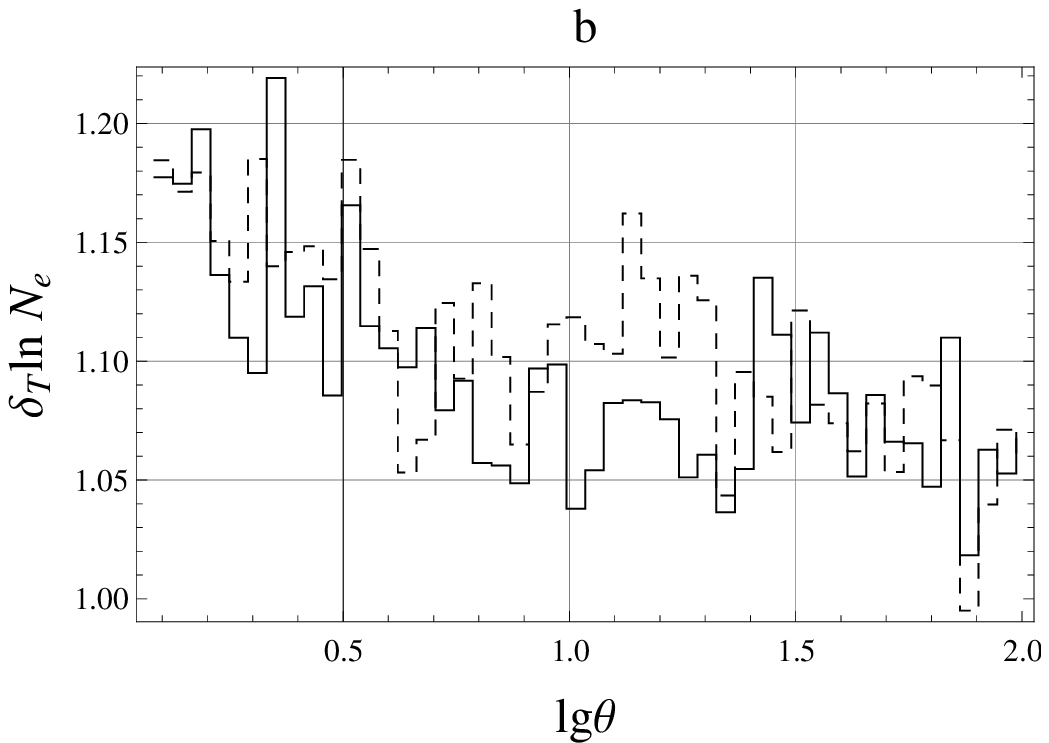}}\\
\end{center}
\caption{Autophasing time (a)  and its RMSD
(b) versus the nonisochronism parameter $\theta$ [black curve --- $N_e=6.75\cdot10^4$ and dashed curve --- $1.08\cdot10^6$].}
\label{Fig.2}
\end{figure}
In numerical analysis  of statistical fluctuations of cooperative
radiation in the presence of shot noise, instead of the number
$N_e$ of real electrons we took the number $N=36\ll N_e$ of
simulated electrons  with the charge $eN_e/N$ and initial phases
\begin{equation}
\phi_k(0)=\frac{2\pi k}{N}+\sqrt{\frac{12N}{N_e}}r_k,k=1..N,
\end{equation}
where $r_k$ are random variables uniformly distributed over the
interval $[0;1)$. It has been shown in  \cite{Penman1992} that
this procedure, boosting the performance of the program, correctly
simulates the shot noise in the absence of energy spread. We
selected the following values of controlling parameters:
$N_e=6.75\cdot10^4,1.08\cdot10^6$,
$\theta=1$--$107$. The numerical experiment with each
$N_e$--$\theta$ pair was repeated one hundred times.

Fig. 1 and 2 show the results of our computation from which we can
draw some very important conclusions  \cite{AnishchenkoBaryshevsky2016}. First, the relative RMSD of
the dimensionless peak radiated power weakly dependent on the number of
particles $N_e$ (Fig.1) decreases as
$\delta_P\approx4.3/\sqrt{N_e}$.
Second, the autophasing time decreasing as the nonisochronism
 parameter $\theta$ is increased depends logarithmically on the
 number of particles $T_0\sim\ln N_e$. %
 Third, $\delta_T$ reduces
 according to the approximate formula $\delta_T=q/\ln N_e$, where
 $q(\theta)\approx1.1$ slowly decreases as $\lg(\theta)$ varies from 0 to 2.
Extrapolating the obtained dependence of  $\delta_P(N_e)$ and
$\delta_T(N_e)$  to the region with large number of particles,
$N_e=10^9$--$10^{12}$, (typical number of electrons in
single-pass generators), we get the estimates
$\delta_T=0.04$--$0.05$ and $\delta_P<10^{-4}$. From this we can
deduce that shot noise leads to a
$4$--$5\%$ fluctuation of the autophasing time at insignificant
fluctuation of the peak radiated power.

Let us note here that phase premodulation of
electrons-oscillators:
\begin{equation}
\label{shot1}
\phi_k(0)=\frac{2\pi
k}{N}+\sqrt{\frac{12N}{N_e}}r_k+\delta_\phi\cos\Big(\frac{2\pi
k}{N}\Big),k=1..N,
\end{equation}
having practically no effect on  $P_0$, $T_0$,  and $\delta_P$,
leads to a noticeable decrease in the fluctuation of autophasing
time: $\delta_T$ no longer decreases logarithmically ($\delta_T\sim1/\ln N_e$),
but as $1/\sqrt{N_e}$ (Fig. 3). In the expression \eqref{shot1}, $\delta_\phi\ll1$ is the premodulation parameter.
\begin{figure}[ht]
\begin{center}
 \resizebox{65mm}{!}{\includegraphics{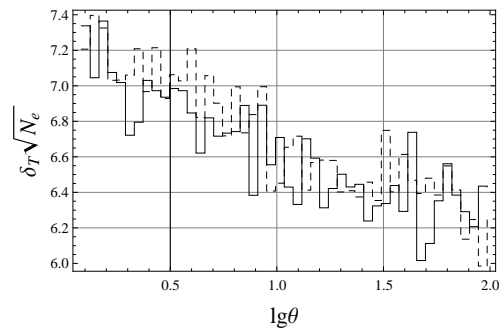}}\\
\end{center}
\caption{RMS fluctuations of autophasing time versus the
nonisochronism parameter $\theta$ at $\delta_\phi=0.05$ [black curve --- $N_e=6.75\cdot10^4$ and dashed curve --- $1.08\cdot10^6$].}
\label{Fig.3}
\end{figure}

\section{Autophasing time}

In the previous section we have shown that  fluctuations in the
cooperative radiation caused by shot noise depend on the degree to
which the electrons-oscillators are premodulated. It has been
found, in particular, that premodulation results in that the
relative RMSD of the autophasing time $\delta_T$ becomes
square-root dependent on $N_e$ ($\delta_T\sim1/\sqrt{N_e}$)
instead of logarithmically dependent ($\delta_T\sim1/\ln N_e$).
Let us show how this important result can be obtained analytically.

According to \eqref{wainshtein1} without premodulation, the
contribution coming from
 each particle   to the average oscillation amplitude $a_{in}=a(0)$ is $a_k(0)=e^{i\phi_k}/N_e$.
The contribution coming from a premodulated particle
$a_k(0)=e^{i\phi_k+i\delta_\phi\cos(\phi_k)}$ (the initial phases
$\phi_k$ are uniformly distributed over the interval $[0;2\pi)$).
By averaging over $\phi_k$, we shall find the average value of
$a_{in}$ as well as the RMSD thereof at $N_e\gg1$ and
$\delta_\phi\ll1$:
 \begin{equation}
\label{phasing1}
\begin{split}
 &<\text{Im}a_{in}>=J_1(\delta_\phi),\\
 &<\text{Re}a_{in}>=0,\\
 &|<\text{Im}a_{in}>^2-<\text{Im}^2a_{in}>|^{1/2}=\sqrt{\frac{1}{2N_e}(1+J_2(2\delta_\phi))}\\
 &\approx\frac{1}{\sqrt{2N_e}},\\
 &|<\text{Re}a_{in}>^2-<\text{Re}^2a_{in}>|^{1/2}=\sqrt{\frac{1}{2N_e}(1-J_2(2\delta_\phi))}\\
 &\approx\frac{1}{\sqrt{2N_e}},\\
\end{split}
\end{equation}
where $J_2(2\delta_\phi)$ is the Bessel function.
\begin{figure}[ht]
\begin{center}
 \resizebox{65mm}{!}{\includegraphics{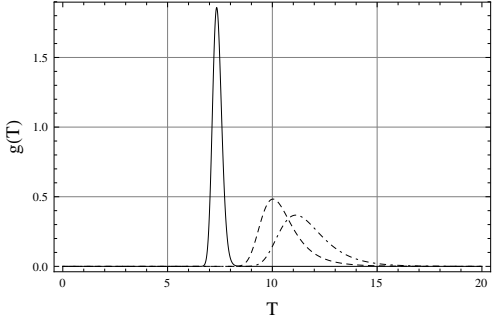}}\\
\end{center}
\caption{Autophasing time distribution functions at
$\tau_0=1$ and $N_e=1.08\cdot10^6$ [solid
curve --- $\delta_\phi=0.05$, dashed curve --- $\delta_\phi=0.01$, dot-dashed
curve --- $\delta_\phi=0$].} \label{Fig.4}
\end{figure}
\begin{figure}[ht]
\begin{center}
 \resizebox{65mm}{!}{\includegraphics{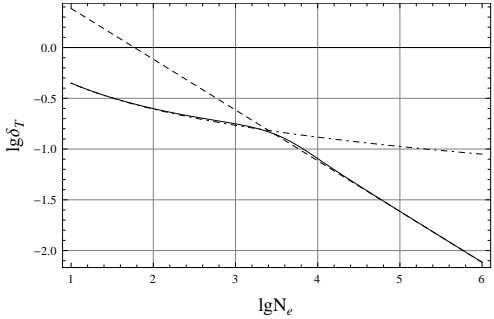}}\\
\end{center}
\caption{RMSD as a function of the number of particles at
 $\delta_\phi=0.05$
[solid curve corresponds to  \eqref{phasing5}, dashed curve
corresponds to the asymptotic expansion at $\delta_\phi^2N_e\gg1$,
dot-dashed curve --- to the asymptotic expansion at
$\delta_\phi^2N_e\ll1$].} \label{Fig.5}
\end{figure}
Since $N_e\gg1$, then in view of the central limiting theorem,
 $\text{Re}a_{in}$ and  $\text{Im}a_{in}$ have a Gaussian
distribution:
\begin{equation}
\label{phasing2}
 f(\text{Re}a_{in},\text{Im}a_{in})=\frac{N_e}{\pi}e^{-N(\text{Re}^2a_{in}+\text{Im}^2a_{in})},
\end{equation}
leading  to that the  radiation power distribution
$P_{in}=2|a_{in}|^2$ at the initial time has the form:
\begin{equation}
\label{phasing3}
 f(P_{in})=N_ee^{-N_eP_{in}/2-N_e\alpha}I_0(N_e\sqrt{2P_{in}\alpha}).
\end{equation}
where  $I_0$ is the modified Bessel function, $\alpha=J_1^2(\delta_\phi)\approx\delta_\phi^2/4$.

Let us estimate the autophasing time by formula
\begin{equation}
\label{phasing4}
 T=\tau_0\ln(P_0/P_{in})=\tau_0\ln(2/P_{in}).
\end{equation}
Here, $\tau_0=1/\text{Re}(-1+\sqrt{1+4i\theta})$
\cite{Wainshtein1990} is the time period needed for  the radiation
power to increase by a factor of $e$ in the linear instability
stage. Equation \eqref{phasing4} implies, first, that the  linear
stage is much longer than the time during which nonlinear effects
are essential, and, second, the radiated power in the saturation
stage is $P_0=2|a|^2\sim2$. The latter condition
means coherent summation of oscillations from all the particles.

Using  \eqref{phasing3} and  \eqref{phasing4}, we can find the
distribution function  $g(T)$ related to  $f(P_{in})$ by
\begin{equation}
\label{phasing5}
 g(T)=f(P_{in})|dP_{in}/dT|.
\end{equation}
Thus we have
\begin{equation}
\label{phasing6}
\begin{split}
 &g(T)=\frac{N}{\tau_0}I_0(2N_e\sqrt{\alpha e^{-T/\tau_0}})\\
 &\times\exp(-N_ee^{-T/\tau_0}-N_e\alpha-T/\tau_0).\\
\end{split}
\end{equation}
From the distribution functions $g(T)$ plotted in  Figure
\ref{Fig.4} follows that with growing premodulation, the
autophasing time and its RMSD thereof decrease.

Using the distribution function $g(T)$, we shall compute the
relative RMSD $\delta_T$:
\begin{equation}
\label{phasing7}
\begin{split}
&\delta_T=\frac{\sqrt{|<T^2>-<T>^2|}}{<T>}\\
&=\frac{\sqrt{\pi^2/6-e^{-2N_e\alpha}F_{-1}^2(N_e\alpha)+e^{-N_e\alpha}G_{-1}(N_e\alpha)}}{e^{N_e\alpha}(\gamma_e+\ln N_e)+F_{-1}(N_e\alpha)},\\
&F_x(y)=\frac{\partial L_x(y)}{\partial x},\\
&G_x(y)=\frac{\partial^2 L_x(y)}{\partial x^2}.\\
\end{split}
\end{equation}
Here,  $\gamma_e=0.577$ is the Euler constant and $L_x(y)$ is the Laguerre polynomial. If  $N_e\alpha\ll1$,
then \eqref{phasing7} transforms to the form
\begin{equation}
\label{phasing8}
\delta_T=\frac{\pi}{\sqrt{6}(\gamma_e+\ln N_e)},
\end{equation}
in the opposite case ($N_e\alpha\gg1$),
\begin{equation}
\label{phasing9}
\delta_T=\frac{\sqrt{2}}{\sqrt{N_e\alpha}\ln(1/\alpha)}.
\end{equation}
The Fig. \ref{Fig.5} plots $\delta_T$ versus
$\lg N_e$, respectively. It is obvious from
\ref{Fig.5} that when the number of particles exceeds a certain
value depending on the degree of premodulation, the logarithmic
$N_e$ dependence of the relative RMSD of the autophasing time
$\delta_T$  ($\delta_T\sim1/\ln N_e$) goes to a square-root
dependence ($\delta_T\sim1/\sqrt{N_e}$). As follows from \eqref{phasing7}, the value of $\delta_T\sqrt{N_e}=7.67$ at
$\delta_\phi=0.05$ and $0<\lg\theta<2$ agree with the
simulated ones (Fig. \ref{Fig.3}) within 20\% accuracy.

Let us note a very important circumstance.
The
duration of the cooperative radiation pulse  from nonisochronous
electrons-oscillators is $\sim\tau_0$ \cite{Wainshtein1990}.  The RMSD of the
autophasing time in the absence of premodulation has
the same order of magnitude: 
\begin{equation}
\label{phasing10}
\Delta T=\sqrt{|<T^2>-<T>^2|}\approx\pi\tau_0/\sqrt{6}
\end{equation} 
at $N_e\alpha\ll1$.
However, phase premodulation of particles leads to an appreciable decrease
in the autophasing time spread: 
\begin{equation}
\label{phasing11}
\Delta T\approx\tau_0\sqrt{2/N_e\alpha}
\end{equation} 
at $N_e\alpha\gg1$. 

In coherent summation of
cooperative-radiation pulses, the oscillation phases of  all
single-pass generators must differ by $\Delta\phi\ll\pi$, thus  posing the following
limitation on the  average statistical spread of autophasing time
 $\Delta t=T_U\Delta T$:
\begin{equation}
\label{sum1}
\Omega T_U\Delta T\ll \pi,
\end{equation}
giving, after the substitution of  $T_U$ \eqref{Wainstein4b} and $\Delta T$ \eqref{phasing11}
\begin{equation}
\label{sum3}
\sqrt{N_e}\delta_\phi\gg\frac{4\sqrt{2}}{\pi}\frac{\theta}{\text{Re}(-1+\sqrt{1+4i\theta})}\frac{c^2}{v_{\perp0}^2}.
\end{equation}
Let  $v/c\sim0.4$ and $\theta=100$, then we have the following
estimate of the required degree of premodulation:
$\sqrt{N_e}\delta_\phi\gg86$. At $N_e=10^9$, the parameter of
premodulation  $\delta_\phi$ must be greater than
$3\cdot10^{-3}$.

\section{Energy spread}
To take  account of the electron energy spread $E_k(0)\approx
a_k^2/N_e$, we assume  the initial amplitudes $a_k(0)$ to be
Gaussian random variables whose mean equals unity and the relative RMSD
$\delta_a\approx\delta_E/2$ ($\delta_E$ is the relative RMSD
deviation of particle energy).

It should be noted the Penman-McNeil algorithm isn't applicable in the presence of energy spread. Therefore instead of the number
$N=36$ of simulated electrons we took the number $N=288\gg1$ of
particles with the initial phases uniformly
distributed in the interval  $[0;2\pi)$.

\begin{figure}[ht]
\begin{center}
 \resizebox{65mm}{!}{\includegraphics{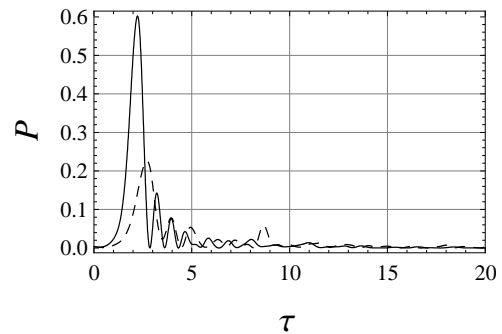}}\\
\end{center}
\caption{Radiated power as a function
of time at $\theta=10$ and different $\delta_a$ [solid
curve --- $\delta_a=0.0$, dashed curve --- $\delta_a=0.04$].}
 \label{Fig.6}
\end{figure}

\begin{figure}[ht]
\begin{center}
 \resizebox{65mm}{!}{\includegraphics{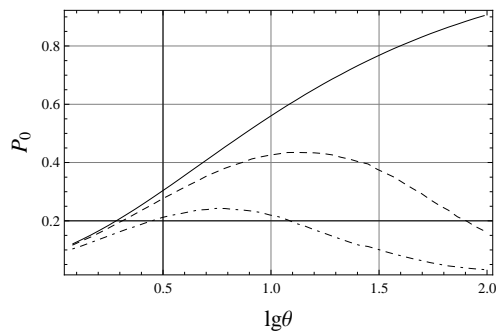}}\\
\end{center}
\caption{Peak radiated power versus the nonisochronism parameter
$\theta$ at different $\delta_a$ [black curve --- $\delta_a=0.00$,
dashed curve --- $\delta_a=0.02$, and
dot-dashed curve --- $\delta_a=0.04$].} \label{Fig.7}
\end{figure}

Analyzing the results of numerical experiments \cite{AnishchenkoBaryshevsky2016}, we can see that
the energy spread leads to a sharp
drop in the radiated power (Fig. \ref{Fig.6}).
This is well-illustrated by
the Fig.  \ref{Fig.7}, where the growing influence of
the energy spread with higher $\theta$ is seen clearly: the energy
spread leads to a stronger suppression of radiation at large
$\theta$, and peak radiated power and radiated energy  both
decrease. 
At $\delta_a=0.02$ ($\delta_E=0.04$), the maximum attainable radiated power
reduces by more than a factor of two.


\section{Conclusion}
In this paper we have studied the statistical properties of
cooperative radiation from an ensemble of nonisochronous
electrons-oscillators interacting with one another via the
radiation field. It has been shown that for the number of
electrons $N_e\sim10^9$--$10^{12}$, typical of modern acceleration
facilities, the relative RMSD  of the autophasing time from its
mean is $\delta_T\approx1.1/\ln N_e\sim0.04$--$0.05$.
The fluctuations of the peak radiated power appear to be
negligibly small $\delta_P<10^{-4}$.

The autophasing time distribution function depending on the
number of particles $N_e$ and the degree of premodulation has
been obtained in the absence of energy spread.
We used this function to show that when the number of particles
exceeds a certain value depending on the degree of premodulation,
the logarithmic dependence of the relative RMSD of the
autophasing time  on  the number of electrons-oscillators
($\delta_T\sim1/\ln N_e$) goes to a square-root dependence
($\delta_T\sim1/\sqrt{N_e}$).

It has also been demonstrated that a slight energy spread
~($\sim4\%$) results in a twofold drop of the maximum attainable
power of cooperative radiation.

The analysis made here indicates
that shot noise, and electron velocity spread pose considerable
constraints on the output characteristics of  single-pass CRMs and
limits the possibility of coherent summation of cooperative
radiation pulses  from several sources.

\appendix
\section{Relativistic electrons-oscillators}
Let in laboratory reference there be  a relativistic electron
beam whose velocity is directed at a small angle to the
magnetic-field vector. The particles in the beam move in a spiral
under Lorentz force $\vec v\times\vec H/c$ .
Let us denote the particle velocity components perpendicular and
parallel to vector $\vec H$ by $\vec u_\perp$ and $\vec
u_z$, respectively. Then the velocity component $\vec
v_\perp$ perpendicular to the magnetic field $\vec H$ in the reference frame attached to the beam is given by
  \cite{Landau2}:
\begin{equation}
 \label{relativistic1}
\vec v_\perp=\frac{\vec u_\perp}{\sqrt{1-\frac{u_z^2}{c^2}}}.
\end{equation}
Let us assume that $v_\perp\ll c$, which leads to the limitations
on  the acceptable value of $u_\perp$:
\begin{equation}
 \label{relativistic2}
\frac{u_\perp}{c}\ll\sqrt{1-u_z^2/c^2}\approx\frac{1}{\gamma}.
\end{equation}

In this section we shall focus on the distribution of the
cooperative radiation from relativistic electrons-oscillators in
the laboratory frame if the condition  \eqref{relativistic2} is
fulfilled.
 In a moving frame of reference, the radiated power  per unit solid angle is given by \cite{Ginzburg,Landau2}
\begin{equation}
\label{relativistic3}
 \frac{dP}{d\Omega}=\frac{3P}{16\pi}(1+\cos^2{\theta}),
\end{equation}
where  $P$ is the total power of electrons-oscillators. The
question naturally arises of  how the radiation intensity
\eqref{relativistic3} can be transformed into the lab reference
frame.

Let us denote the radiated power  per unit solid angle in the lab
frame by $dP_u/d\Omega_u$. Before  seeking $dP_u/d\Omega_u$, let
us note that  the quantity
\begin{equation}
\label{relativistic4}
 X_1=\frac{dP}{d\Omega}\frac{d\Omega dt}{\hbar\omega}
\end{equation}
that equals the number of quanta emitted in solid angle $d\Omega$
during the time  $dt$ is the Lorentz invariant, and  $X_2$ is the
Lorentz invariant, too \cite{Landau2}
\begin{equation}
\label{relativistic5}
 X_2=\frac{d^3k}{\hbar\omega}=\frac{k^2dkd\Omega}{\hbar\omega}.
\end{equation}
The division of  $X_1$ by $X_2$ gives another invariant
\begin{equation}
\label{relativistic6}
X_3=\frac{dt}{k^2dk}\frac{dP}{d\Omega}.
\end{equation}
By equating the invariants in both frames of reference we get
\begin{equation}
\label{relativistic7}
\frac{dP_u}{d\Omega_u}=\frac{k^2_udk_udt}{k^2dkdt_u}\frac{dP}{d\Omega}.
\end{equation}
Because the absolute values of the wave vectors in  moving ($k$)
and fixed ($k_u$) frames of reference are related as
\cite{Landau2}
\begin{equation}
\label{relativistic8}
 k=k_u\frac{1-\frac{u_z\cos(\theta_u)}{c}}{\sqrt{1-\frac{u_z^2}{c}}},
\end{equation}
where $\theta_u$ is the angle between vectors $\vec H$ and $\vec
k_u$, whereas
\begin{equation}
\label{relativistic9}
 \frac{dt}{dt_u}=\frac{1}{\sqrt{1-u_z^2/c^2}}\approx\gamma,
\end{equation}
we can find
\begin{equation}
\label{relativistic10}
\begin{split}
 &\frac{dP_u}{d\Omega_u}=\Bigg(1+\Big(\frac{\cos\theta_u-u_z/c}{1-u_z\cos\theta_u/c}\Big)^2\Bigg)\\
 &\times\frac{(1-u_z^2/c^2)^2}{(1-u_z\cos\theta_u/c)^3}\frac{3P}{16\pi}.\\
\end{split}
\end{equation}

According to  \cite{Landau2,Ginzburg}, the angular distribution of
the power $dP_{det}/d\Omega_u$ detected by the
resting detector in the lab frame is related to
 $dP_{u}/d\Omega_u$ as
\begin{equation}
\label{relativistic11}
 \frac{dP_{det}}{d\Omega_u}=\frac{1}{1-u_z\cos\theta_u/c}\frac{dP_u}{d\Omega_u}.
\end{equation}
Integration of \eqref{relativistic11} with due account of
\eqref{relativistic10} finally gives  ($\gamma\gg1$)
\begin{equation}
\label{relativistic12}
 P_{det}=\frac{5+2u_z^2}{5-5u_z^2}P\approx\frac{7\gamma^2P}{5}.
\end{equation}

\section{Effects of Coulomb and induction fields}
 Let us study the applicability boundaries of
 \eqref{oscillators6}. The equation set  \eqref{oscillators6} was
 derived in the assumption that  the beam size $R_0$ is
 much less than the radiation wavelength
\begin{equation}
 \label{limit1}
R_0\ll\lambda=\frac{mc^2}{eH}.
\end{equation}
We also neglected the Coulomb repulsion and induction fields.
The latter assumption is often
taken in solving specific problems \cite{Wainshtein1990}, so we
shall qualitatively outline its legitimacy.

It has been shown in  \cite{Wainshtein1990,Ilyinsky1988} that the
possibility of phasing electrons-oscillators without the external
field arises because the particles oscillate nonisochronousally,
i.e., the particle oscillation frequency depends on energy (see
\eqref{oscillators3}). As a result, a nonlinear term having the
same order of magnitude as $\Omega v^3/c^2$ appears in
\eqref{oscillators6}. The Coulomb and induction fields have no
effect of autophasing if the acceleration $\sim
\frac{Ne^2}{mR_0^2}$ induced to the electron by the Coulomb forces
(induction fields are by a factor of $c/v$ less than the Coulomb
ones) is much smaller than $\Omega v^3/c^2=\frac{eHv^3}{c^3}$. Let
us write this condition in the form:
\begin{equation}
 \label{limit2}
R_0\gg R_{min}=\sqrt{\frac{N_ee c^3}{Hv^3}}.
\end{equation}

Let $H=6$~kGs, $v/c=0.4$, and  $N_e=10^{9}$. Then
$\lambda=0.3$~cm, $R_{min}=0.04$~cm, and the particle orbit
radius $R_{orbit}=v/\Omega=mcv/eH=0.1$~cm is between  $R_{min}$
and $\lambda$. That is why if in this specific case the beam size
is the same as the particle orbit, we can satisfy both
\eqref{limit1} and \eqref{limit2} requirements.

\section{Near-field of charged particles}
Let us consider a short beam of charged particles moving with
nonrelativistic velocities. The beam particles produce the
electromagnetic field that can be described using the scalar
$\phi$ and vector  $\vec A$ retarded potentials \cite{Landau2}:
\begin{equation}
 \label{NF1}
\begin{split}
\phi(\vec r,t)=\int\frac{1}{R}\rho(\vec r_1,t-R/c)dV_1,\\
\vec A(\vec r,t)=\int\frac{1}{R}\vec j(\vec r_1,t-R/c)dV_1,\\
\end{split}
\end{equation}
where $R=|\vec r-\vec r_1|$ is the distance from the volume
element  $dV_1$ to the point of observation $\vec r$, where we
seek the potential difference, $\rho$ and $\vec j$ are the current
and charge densities, respectively. The electric  $\vec E$ and
magnetic  $\vec H$ fields  relate to the potentials  $\phi$ and
$\vec A$ as
\begin{equation}
 \label{NF2}
\begin{split}
\vec E=-\frac{1}{c}\frac{\partial A}{\partial t}-\nabla\cdot\phi,\\
\vec H=\nabla\times\vec A.\\
\end{split}
\end{equation}
Using these relation, we can readily find the force acting on each
charge from all beam particles
\begin{equation}
 \label{NF3}
\vec F=e(\vec E+\vec v\times \vec H/c).
\end{equation}
Because the velocities of all charges  are small compared to the
speed of light, their distribution hardly can change a lot during
the time $\sim R_0/c$ ($R_0$ is the size of the
beam).
That is why, to find forces acting on the charges, we can expand
$\rho$ and  $\vec j$ into power series in $R/c$. For the
scalar potential, we find accurate up to the third order in terms
of  ~$c^{-1}$:
\begin{equation}
 \label{NF4}
 \begin{split}
&\phi(\vec r,t)=\int\frac{\rho}{R}dV-\frac{1}{c}\frac{\partial}{\partial t}\int\rho dV\\
&+\frac{1}{2c^2}\frac{\partial^2}{\partial t^2}\int R\rho dV-\frac{1}{6c^3}\frac{\partial^3}{\partial t^3}\int R^2\rho dV,\\
\end{split}
\end{equation}
But $\int\rho dV$ is the constant total charge of the system; Then
the second term in this expression equals zero, and so
\begin{equation}
 \label{NF4b}
 \begin{split}
&\phi(\vec r,t)=\int\frac{\rho}{R}dV+\frac{1}{2c^2}\frac{\partial^2}{\partial t^2}\int R\rho dV\\
&-\frac{1}{6c^3}\frac{\partial^3}{\partial t^3}\int R^2\rho dV.\\
\end{split}
\end{equation}
We can do the same with $\vec A$. But $c^{-1}$ appearing in the
expression for the vector potential is multiplied by $c^{-1}$ when
$\vec A$ is substituted into the expression for the force. Because
we seek the potentials within the  accuracy up to the third order
in terms of ~$c^{-1}$, in the expansion of $\vec A$ suffice it to
find the first two terms, i.e.,
\begin{equation}
 \label{NF5}
\vec A(\vec r,t)=\frac{1}{c}\int\frac{\vec j}{R}dV-\frac{1}{c^2}\frac{\partial}{\partial t}\int\vec j dV,
\end{equation}

Let us perform gauge transformation of the potentials:
\begin{equation}
 \label{NF6}
\begin{split}
\phi_{new}=\phi-\frac{1}{c}\frac{\partial f}{\partial t},\\
\vec A_{new}=\vec A+\nabla f,\\
\end{split}
\end{equation}
choosing the function  $f$ such that the scalar potential
$\phi_{new}$ becomes
\begin{equation}
 \label{NF7}
\phi_{new}=\int\frac{\rho}{R}dV:
\end{equation}
\begin{equation}
 \label{NF8}
f=\frac{1}{2c}\frac{\partial}{\partial t}\int R\rho dV-\frac{1}{c^2}\frac{\partial^2}{\partial t^2}\int R^2\rho dV.
\end{equation}
Then a new vector potential is
\begin{equation}
 \label{NF9}
\begin{split}
&\vec A_{new}=\frac{1}{2c}\nabla\Big(\frac{\partial}{\partial t}\int R\rho dV\Big)-\frac{1}{c^2}\frac{\partial}{\partial t}\int \vec j dV\\
&-\frac{1}{3c^2}\frac{\partial^2}{\partial t^2}\int\vec R\rho dV.\\
\end{split}
\end{equation}
Passing from integration to summation over individual charges, we
get  \cite{Landau2}
\begin{equation}
 \label{NF10}
\begin{split}
&\vec A_{new}=\vec A_1+\vec A_2,\\
&\vec A_1=\sum_je\frac{\vec v_j+(\vec v_j\vec n)\vec n}{2cR_j},\\
&\vec A_2=-\sum_j\frac{2}{3c^2}e\dot{\vec v}_j\\
\end{split}
\end{equation}
and
\begin{equation}
 \label{NF11}
\phi_{new}=\sum_j\frac{e}{R_j},
\end{equation}
where $\vec n_j=\vec R_j/R_j$, $\vec R_j=|\vec r-\vec r_j|$.

The first term  $\vec A_1$ in  \eqref{NF10} determines the
induction field of the system of charges and the scalar potential
$\vec A$ determines  the electric field of the spatial charge.
Both fields are non-uniform and depend on the averaged charge
distribution in the electron beam \cite{Wainshtein1990}. The
second term $\vec A_2$ in \eqref{NF10} is uniform and arises
through radiation; the related electric field is given by
\begin{equation}
 \label{NF12}
\vec E_{rad}=\sum_j\frac{2}{3c^2}e\ddot{\vec v}_j.
\end{equation}

\end{document}